\documentclass[12pt]{article}
\usepackage{amsmath}
\usepackage{epsfig}
\usepackage{a4wide}

\begin{document}
\title{Contact forces in regular 3D granular pile}
\author{G. Oron and H.J. Herrmann\\ 
\normalsize Laboratoire de Physique et M\'ecanique des Milieux
H\'et\'erog\`enes,\\
\normalsize UMR CNRS 7636, \'Ecole de Physique et de Chimie Industrielles\\ 
\normalsize de la ville de Paris, \\
\normalsize 10 rue Vauquelin, 75231 Paris Cedex 05, FRANCE \\
\normalsize email : oron@pmmh.espci.fr
  }
\date{september 28 1998}

\maketitle

\begin{abstract}
  We present exact results for the contact
  forces in a three dimensional static piling of identical, stiff and
  frictionless spheres. The pile studied is a pyramid of equilateral
  triangular base (``stack of cannonballs'') with a FCC (face centered
  cubic) structure. We show in particular that, as for the two
  dimensional case, the pressure on the base of such a pile is
  uniform.
\end{abstract}

\paragraph{PAC numbers:}
\begin{description}
\item{83.70.Fn:} Granular solids
\item{46.10.+z:} Mechanics of discrete systems
\end{description}


The force propagation through a heap of granular materials is a
controversial issue~(see refs.~\cite{savage97,cargese-savage} and
references herein) and still attracts much attention. Experiments in
conical sand piles show the existence of a depression in the normal
force on the base (a {\em dip}) under the pile's apex rather than the
intuitively expected maximum~\cite{smid81,huntley97}. Much work has
been done in this subject in order to explain this effect numerically
and theoretically . Most of the results obtained in this domain are
for models considering two dimensional (2D) piles of discs or
rods~\cite{cargese-oron,oron98,luding97,claudin97,clement97,edwards96b,liffman92,liffman94}.
It is unclear however, whether in general, these results can be
directly generalized to three dimensions (3D). In the case of the
continuum approach proposed in ref.~\cite{wittmer97} it was noticed
that the choice of the second closure equation needed for the 3D case
did not have an important impact on the results obtained, so that the
2D results survive the passage from 2 to 3 dimensions. This case is
a priori unlikely when discrete models are considered.  One
difficulty might be that a simple 3D extension of a 2D pile, which is
obtained by a revolution around the symmetry axis of a 2D pile is not
any regular 
3D piling of spheres but rather a 3D piling of tori. Another
possible problem is that in 3D there are more packing possibilities
than in 2D. Even if one accepts that each layer of spheres is arranged
on a triangular lattice, there are different ways to pile layers one
on top of the other, giving different arrangements; FCC, HCP, DHCP
etc.  Another issue is the shape of the base. A circular base,
as for the real conical sandpile, is ``unnatural'' for spheres in the
3D context, while in 2D it is the simplest one since for a 2D pile,
a circular base is simply a segment.

\begin{figure}[p]
  $$
  \epsfig{file=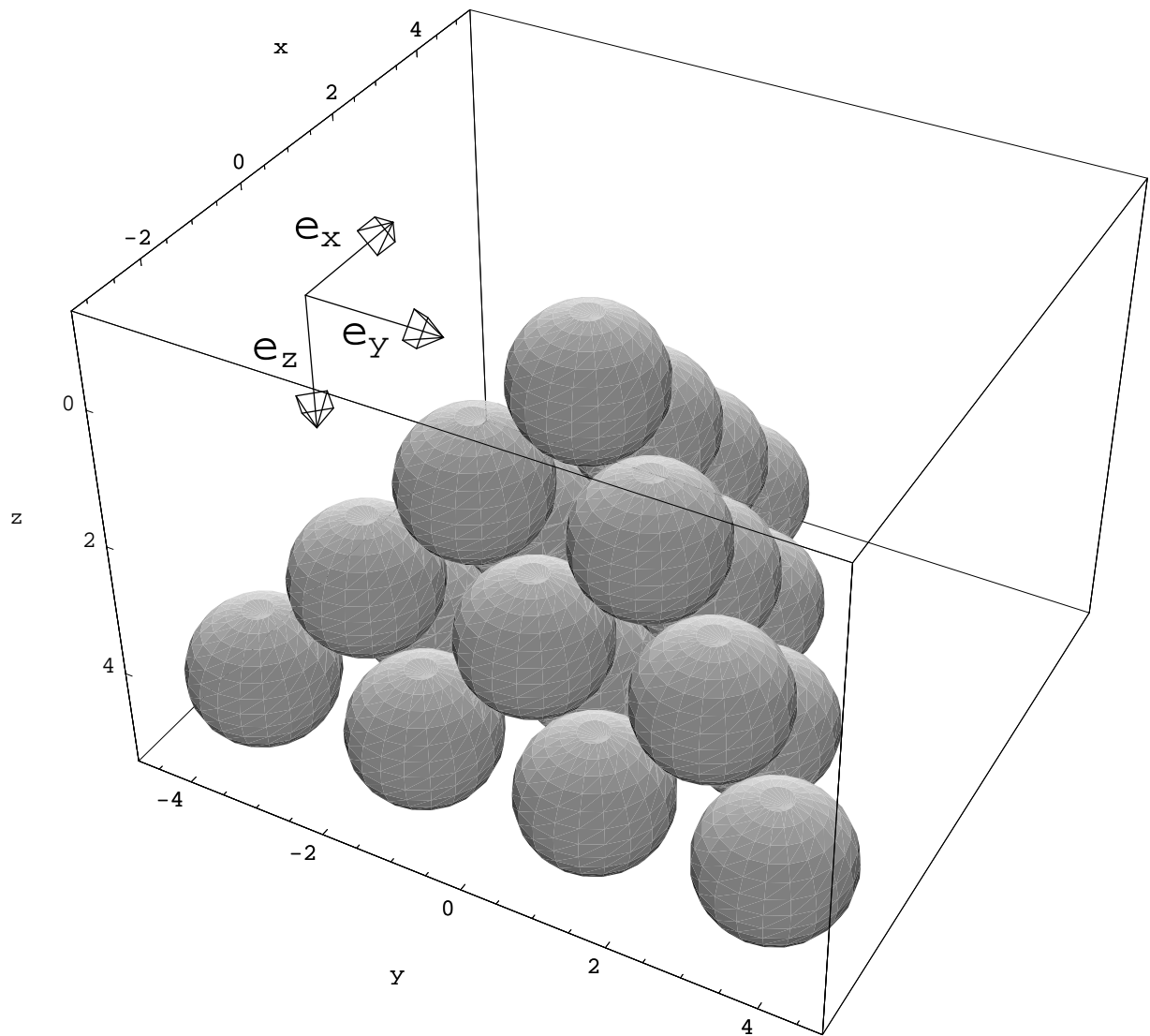,width=\linewidth}
  $$
  \caption{A pile of four layers with $a=1.2\ R$.} 
  \label{fig:pile}
\end{figure}

Here we study a 3D case that might be considered as an extension of
the one proposed in refs.~\cite{hong93,liffman94} for 2D.  The
configuration we study is presented in fig.~\ref{fig:pile}.  It is a
regular 3D pyramid with triangular base and an FCC-like
structure\footnote{the structure is exactly FCC if $a=R$, in which
  case the diagonal of the elementary cube is oriented vertically},
composed of identical, spherical, stiff and frictionless particles of
mass $m$ and radii $R=1$ (this piling is also known as the ``spanish
stack of cannonballs'')~\cite{ashcroft}.  In each horizontal layer the
distance between neighboring spheres is $2a$, where $1< a< 3/2$ so
that there is no overlap between spheres, horizontally or vertically.
Our aim is to calculate the contact forces everywhere in the pile,
especially the normal and shear forces applied by the pile on the
supporting surface.  For the reduced 2D case it was
shown~\cite{hong93} that the pressure profile on the supporting
surface is uniform.  This feature was also proved in
ref.~\cite{huntley93} for a similar pile with periodic vacancies.

\begin{figure}[p]
  $$
  \epsfig{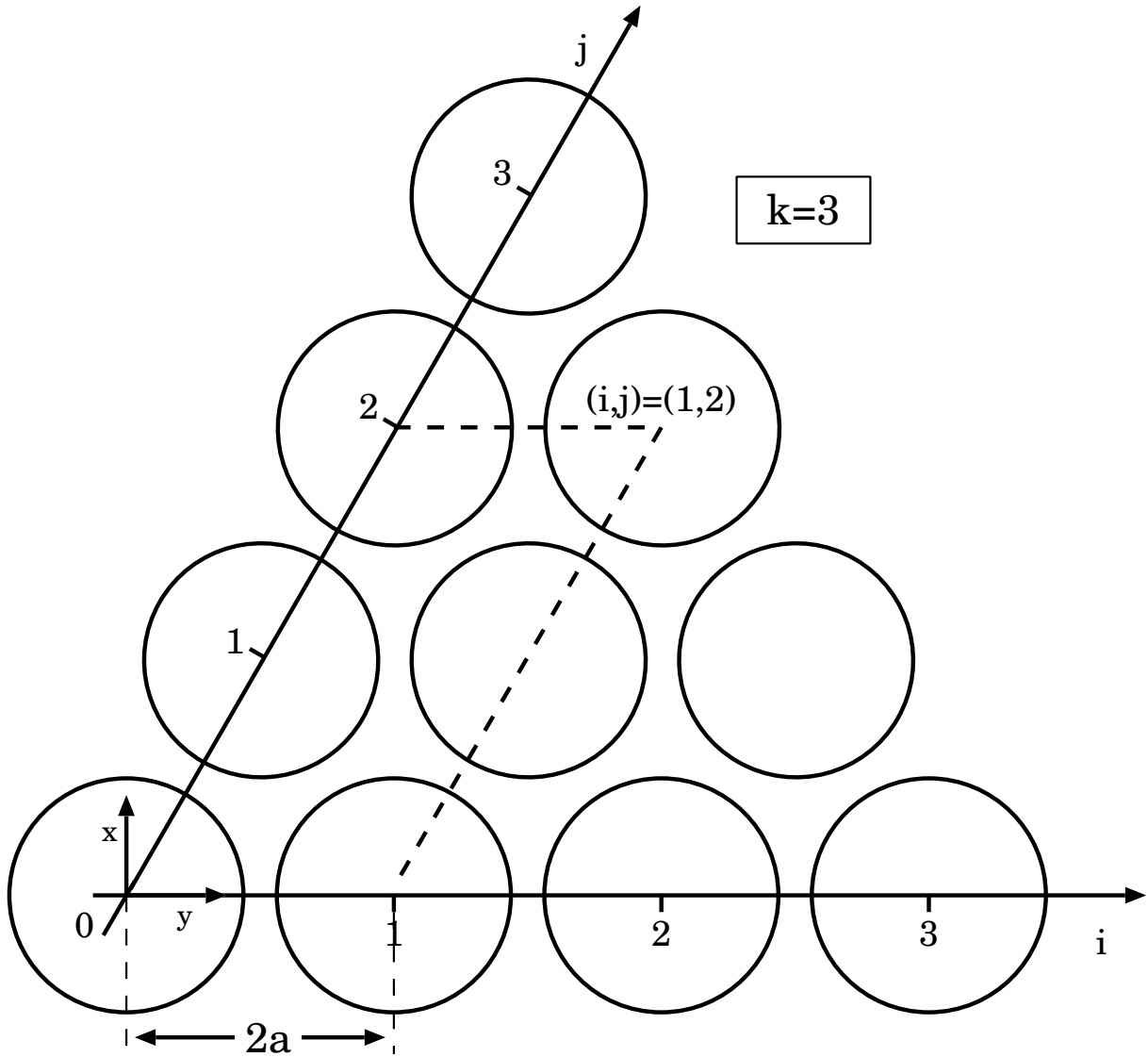}
  $$
  \caption{Indexing convention of the spheres for a layer in the
    pile. The layer shown is the one for $k=3$ (forth from the top).
    Remark that force components are given in the vectorial
    base shown in fig.~\ref{fig:pile} and not in the one used for
    indexing.}
  \label{fig:indexing}
\end{figure}

In the following, we will index the spheres using a triplet of
integers $(i,j,k)$ where $k$ is the horizontal layer index counted
from the top of the pile down (the topmost layer has $k=0$) and the
couple $(i,j)$ indices the spheres in the $k$-th layer as shown in
fig.~\ref{fig:indexing}.
If the pile has a total of $L$ layers then, $0\le k\le L-1,\ 0\le j\le
k$ and~$\ 0\le i\le k-j$.  Components of the force vectors will be
given in the vectorial base shown in fig.~\ref{fig:pile}.

\begin{figure}[p]
  \epsfig{file=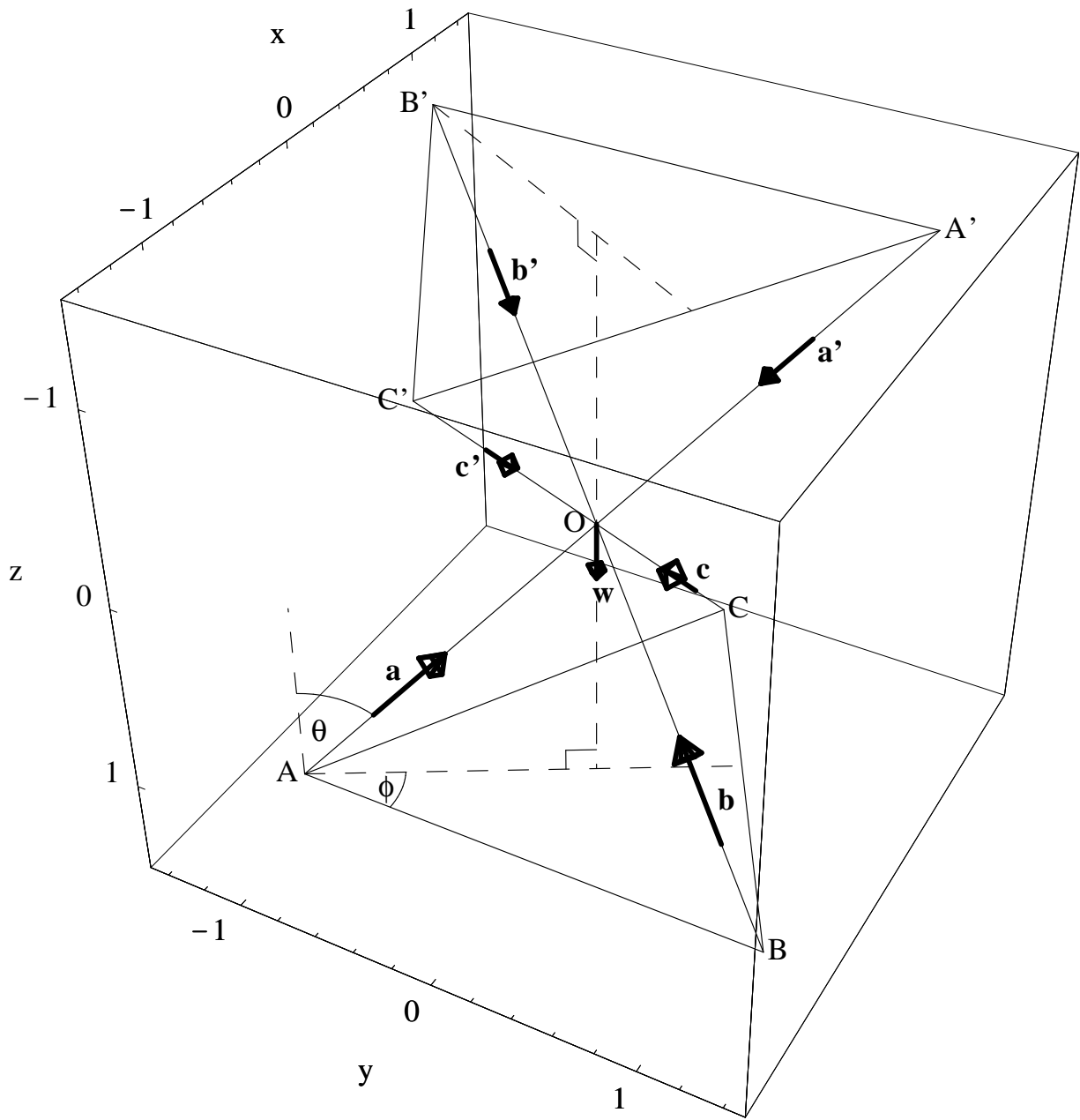,width=\linewidth}
  \caption{A schematic drawing of the forces acting on a given sphere
    centered at $O$ (not shown). The neighboring spheres are the ones
    centered at $A,B$ and~$C$ (layer beneath) which apply the contact
    forces ${\bf a},{\bf b}$ and~${\bf c}$ respectively, and $A',B'$
    and~$C'$ (layer above). which apply the contact forces ${\bf
      a}',{\bf b}'$ and~${\bf c}'$ respectively. $w$ is the weight of
    the particle.}
  \label{fig:forces}
\end{figure}

Let us take a closer look at one of the spheres, indexed: $(i,j,k)$.
We denote, as shown in fig.~\ref{fig:forces}, by ${\bf a},{\bf b}$
and~${\bf c}$ the forces applied by the downwards neighboring spheres,
and by ${\bf a}',{\bf b}'$ and~${\bf c}'$ the forces applied by the
upward spheres. The weight of the particle is ${\bf w}$. One can
easily conclude from fig.~\ref{fig:forces} that
\begin{subequations}
  \begin{eqnarray}
    {\bf a}=\alpha {\bf\hat e}_a, &  {\bf a}'=-\alpha' {\bf\hat e}_a &
    \hbox{with}\quad
    {\bf\hat e}_a=(\sin\theta \sin\phi, \sin\theta \cos\phi, -\cos\theta),
    \label{eq:def-alpha} \\
    {\bf b}=\beta {\bf\hat e}_b, & {\bf b}'=-\beta' {\bf\hat e}_b &
    \hbox{with}\quad
    {\bf\hat e}_b=(\sin\theta \sin\phi,-\sin\theta \cos\phi, -\cos\theta),
    \label{eq:def-beta} \\
    {\bf c}=\gamma {\bf\hat e}_c, & {\bf c}'=-\gamma' {\bf\hat e}_c &
    \hbox{with}\quad
    {\bf\hat e}_c=(-\sin\theta,0,-\cos\theta),   
    \label{eq:def-gamma} 
\end{eqnarray}  
\label{eq:def-abc}
\end{subequations}

\noindent where $\alpha,\beta,\gamma,\alpha',\beta',\gamma'$ are the norms of
the vectors ${\bf a,b,c,a',b',c'}$ respectively, and
${\bf\hat e_a,\hat e_b,\hat e_c}$ are the unit vectors along the the
vectors ${\bf a,b,c}$ respectively.

Since the pile is on top of an horizontal surface the weight of each
sphere is give by
\begin{equation}
  {\bf w} = mg (0,0,1)
  \label{eq:def-weight}.
\end{equation}

The equilibrium of each sphere is given by
\begin{equation}
  \sum {\bf F} = {\bf a}+{\bf b}+{\bf c}+{\bf w} ={\bf 0}.
  \label{eq:equilibrium}
\end{equation}

Combining eqs.~\ref{eq:def-abc},\ref{eq:def-weight} and
\ref{eq:equilibrium} and simplifying we get the following linear
system:

\begin{subequations}
\begin{eqnarray}
  \alpha\sin\phi-\alpha'\sin\phi+ \beta\sin\phi- \beta'\sin\phi
  -\gamma +\gamma' & = & 0 \\
  \alpha- \alpha' - \beta+ \beta' & = & 0 \\
  \alpha\cos\theta -\alpha'\cos\theta+\beta\cos\theta
  -\beta'\cos\theta+\gamma\cos\theta -\gamma'\cos\theta & = & mg
\end{eqnarray}
\end{subequations}

Giving finally
\begin{subequations}
  \begin{eqnarray}
    \Delta\alpha + \Delta\beta - \Delta\gamma / \sin\phi & = & 0 \\
    \Delta\alpha - \Delta\beta & = & 0 \\
    \Delta\alpha + \Delta\beta + \Delta\gamma & = & mg / \cos\theta ,
  \end{eqnarray}
\end{subequations}

\noindent where $\Delta\alpha=\alpha-\alpha',\ \Delta\beta=\beta-\beta'$ and
$\Delta\gamma = \gamma-\gamma'$. This system is easily solved
yielding

\begin{equation}
  \Delta\alpha=\Delta\beta=\frac{mg}{2(1+\sin\phi)\cos\theta}
  \qquad\hbox{and}\quad \Delta\gamma =
  \frac{\sin\phi}{1+\sin\phi}\frac{mg}{\cos\theta}.
\end{equation}

In the case of an equilateral triangular base we have $\phi=\pi/6$, so
that

\begin{equation}
   \Delta\alpha=\Delta\beta=\Delta\gamma=\frac{mg}{3\cos\theta},
\end{equation}

In order to get the force acting on a specific
contact one should count the number of spheres up to the free surface
in the corresponding direction, each one of these spheres
contributing $mg/(3\cos\theta)$. We denote $N_\alpha(i,j,k)$,
$N_\beta(i,j,k)$ and~$N_\gamma(i,j,k)$ the number of particles on top
of the $(i,j,k)$ sphere in the
corresponding direction, so that

\begin{subequations}
  \begin{eqnarray}
    \alpha'(i,j,k) & = & mg/(3\cos\theta)\ N_\alpha(i,j,k), \\
    \beta'(i,j,k) & = & mg/(3\cos\theta)\ N_\beta(i,j,k), \\
    \gamma'(i,j,k) & = & mg/(3\cos\theta)\ N_\gamma(i,j,k).
    \label{eq:forces1}
  \end{eqnarray}
\end{subequations}

One can easily show that (see appendix for details)
\begin{equation}
   N_\alpha(i,j,k)=k-i-j,\quad N_\beta(i,j,k)=i \quad
   \hbox{and} \quad
  N_\gamma(i,j,k)=j,
  \label{eq:counts}
\end{equation}

\noindent yielding, when introduced into eq.~\ref{eq:forces1}
\begin{subequations}
  \begin{eqnarray}
    \alpha'(i,j,k) &=& (k-i-j) mg/(3\cos\theta) \\
    \beta'(i,j,k) &=& i m g/(3\cos\theta)\\
    \gamma'(i,j,k) &=& j m g/(3\cos\theta),
  \end{eqnarray}
  \label{eq:forces2}
\end{subequations}

\noindent which gives us the amplitude of the contact forces anywhere
in the pile. If we suppose that the pile is $L$ layers high and that
the bottom layer spheres are ``glued'' to a flat surface, the force
applied by the bottom layer spheres on the surface; ${\bf
  F}_s(i,j)$ can easily be calculated, since from the equilibrium of a
bottom layer particle we get
\begin{equation}
{\bf F}_s(i,j)={\bf a'}(i,j,L-1)+{\bf b'}(i,j,L-1)+{\bf c'}(i,j,L-1)+
{\bf w},
\end{equation}

and using eqs.~\ref{eq:def-abc},\ref{eq:def-weight}
and~\ref{eq:forces2} in this last equation gives

\begin{eqnarray}
\nonumber  {\bf F}_s(i,j) & = & \frac{mg}{3\cos\theta} \biggl( -1/2\ (L-1-3j)
  \sin\theta, -\sqrt{3}/2\ (L-1-2 i-j) \sin\theta,\\
  & & (L+2)\cos\theta\biggr).
\end{eqnarray}

Thus, the normal force on the base of the pile applied by each one of
the spheres is the same and has an amplitude of
\begin{equation}
  mg(L+2)/3.
  \label{eq:result-normal}
\end{equation}

This force is independent of $\theta$ and hence, of $a$.
One can easily verify that the sum of the normal forces on the surface
is equal to the total weight of the pile $mg\ L(L+1)(L+2)/6$.

\begin{figure}[p]
  \begin{center}
    \epsfig{file=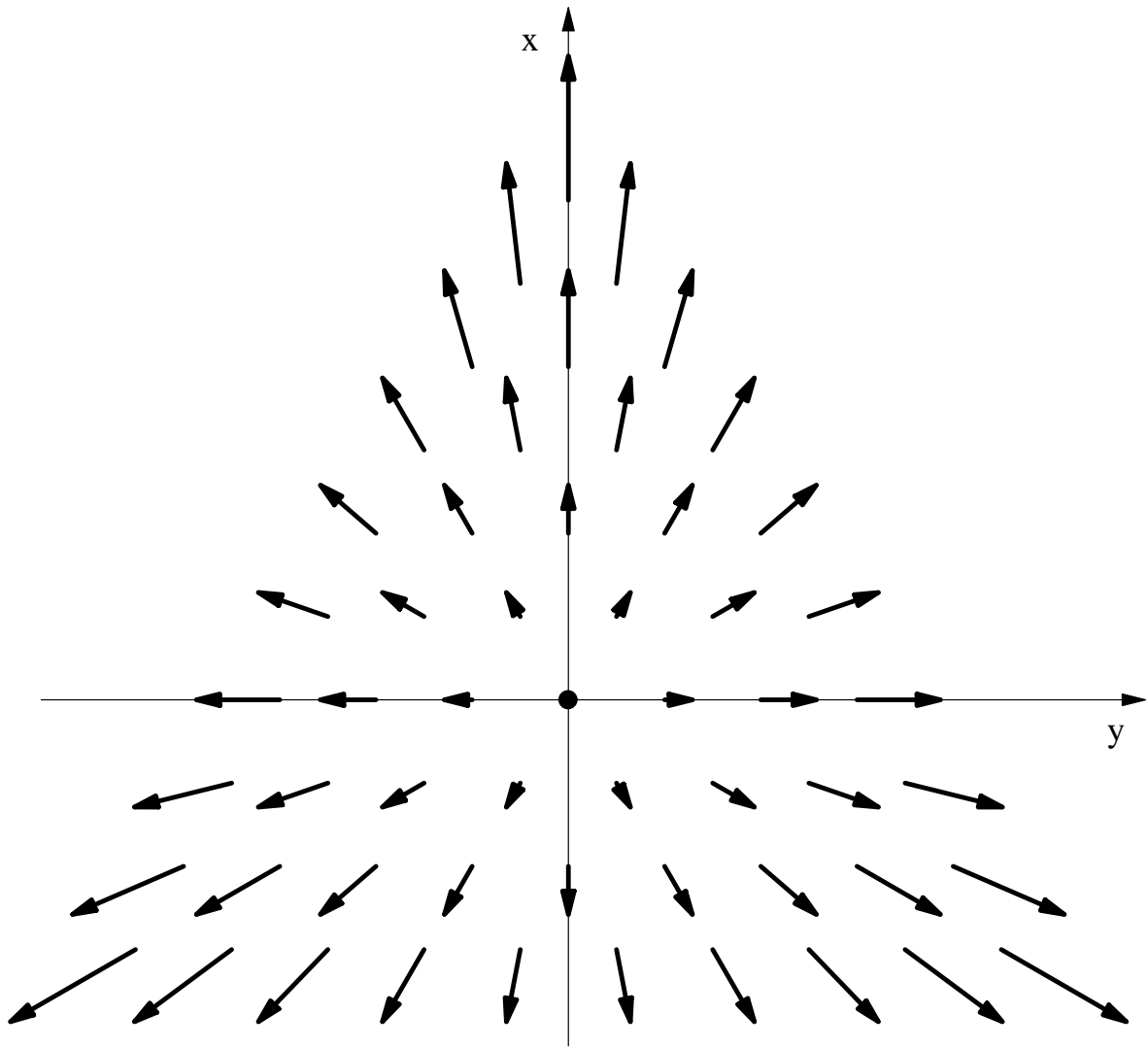,width=\linewidth}
    \caption{A plot of the shear force applied by the pile on the
      supporting surface for a 10 layer pile. Notice that the shear
      force vanishes under the apex of the pile as required by the
      symmetry of the pile.}
    \label{fig:shear}
  \end{center}
\end{figure}

The shear force applied on the surface is given by
\begin{equation}
  \frac{-mg\tan\theta}{3} \left( (L-1-3j)/2,\ \sqrt{3}/2\ (L-1-2 i-j)
  \right).
  \label{eq:result-shear}
\end{equation}
An example of this vector field is given in fig.~\ref{fig:shear}.
Notice that the shear force vanishes under the apex of the pile as
required from symmetry.

In conclusion, in the case of a FCC-like piling with equilateral,
triangular base of stiff, identical, spherical particles we were able
to calculate the contact forces between any couple of neighboring
particles (eq.~\ref{eq:forces2}). In particular we have shown that
normal forces on the base are constant and are given by
eq.~\ref{eq:result-normal} while the shear force is given by
eq.~\ref{eq:result-shear}. These results are similar to those obtained
in ref.~\cite{hong93} for the 2D case and might suggest some
applicability of 2D model results to 3D or, at least, that the
starting point in 3D is the same as for 2D. It is clear, however, that
horizontal contacts should be considered in order to get more physical
behavior like arching. But in this case, the system is hyperstatic and
more complicated techniques are to be used,
exactly~\cite{oron98,cargese-oron} or numerically~\cite{luding97}, in
order to get the correct contact network.  Further investigations
should be carried out in order to obtain similar results for other
base shapes like hexagons and other structures, especially in the RHCP
(Random Hexagonal Compact Packing) case that can be a convenient way
to introduce disorder in the system so to approach a modelization of
real granular piles.

\section{Appendix}
Let us start by showing by induction over $k$ that :
$N_\gamma(i,j,k)=j\quad \forall \{0\le j\le k | 0\le i\le k-j\}$.
When $k=0$ it is clear that $N_\gamma(0,0,0)=0$ since the top particle
does not have any neighbor on it. Now let us suppose that
$N_\gamma(i,j,k)=j\quad \forall \{0\le j\le k | 0\le i\le k-j\}$ for a
certain $k$ and prove it for $k+1$. If $j=0$ the particle is on the
surface of the pile and $N_\gamma(i,0,k+1)=0=j$. When $j>0$ the
$(i,j,k+1)$ sphere is in contact with the $(i,j-1,k)$ particle, so if
$(i,j,k+1)$ particle is centered at point $O$ in
fig.~\ref{fig:forces}, the $(i,j-1,k)$ particle is centered at $C'$
and the contact point is at the middle of the segment
$\overline{OC'}$. Thus, we conclude, using the hypothesis, that
$N_\gamma(i,j,k+1)=N_\gamma(i,j-1,k)+1=j$, Q.E.D.

$N_\alpha(i,j,k)$ and $N_\beta(i,j,k)$ are obtained by a rotation of
the pile around the $z$ axis of $2\pi/3$ and $4\pi/3$ respectively,
which correspond to applying the transformation $\{j\rightarrow
k-i-j,i\rightarrow j\}$ once and twice respectively. 

\bibliographystyle{unsrt} 
\bibliography{pile3d}
\end{document}